\documentclass[osajnl,twocolumn,showpacs,10pt]{revtex4-1} 

\usepackage{amsmath}
\usepackage{amssymb}
\usepackage{graphicx}

\newcommand{\appropto}{\mathrel{\vcenter{
  \offinterlineskip\halign{\hfil$##$\cr
    \propto\cr\noalign{\kern2pt}\sim\cr\noalign{\kern-2pt}}}}}

\begin{document}

\title{Propagation-invariant beams with quantum pendulum spectra: \\ from Bessel beams to Gaussian beam-beams}

\author{Mark R Dennis and James D Ring}

\affiliation{H H Wills Physics Laboratory, University of Bristol, Tyndall Avenue, Bristol BS8 1TL, UK}

\begin{abstract}
We describe a new class of propagation-invariant light beams with Fourier transform given by an eigenfunction of the quantum mechanical pendulum.
These beams, whose spectra (restricted to a circle) are doubly-periodic Mathieu functions in azimuth, depend on a field strength parameter.
When the parameter is zero, pendulum beams are Bessel beams, and as the parameter approaches infinity, they resemble transversely propagating one-dimensional Gaussian wavepackets (Gaussian beam-beams).
Pendulum beams are the eigenfunctions of an operator which interpolates between the squared angular momentum operator and the linear momentum operator.
The analysis reveals connections with Mathieu beams, and insight into the paraxial approximation.
\end{abstract}




\maketitle  

The appreciation of special functions has become a major component of the modern study of structured light beams.
Beams with transverse amplitude distributions specified by various special function profiles given by Bessel functions \cite{durnin}, Airy functions \cite{berrybalazs,Siviloglou2007}, Mathieu functions \cite{MathieuBeams}, and Gaussians times Hermite or Laguerre polynomials \cite{Siegman}, etc., exhibit many interesting physical properties, such as orbital angular momentum (OAM) \cite{absw:laguerre,fap:oam}, propagation invariance \cite{mjgd:spread} and self-healing \cite{BesselRecon}.

Special properties of the functions are often related to the physical problems in which they are typically studied; for instance, higher-order Gaussian beams in the focal plane arise as eigenfunctions of the two-dimensional harmonic oscillator \cite{da:analogies}, both as Hermite-Gauss (HG) modes (quantized back-and-forth harmonic motion), and Laguerre-Gauss (LG) modes (quantized circular harmonic motion) \cite{absw:laguerre}.
The quantum states corresponding to elliptic vibration states are the `Generalized Hermite-Laguerre Gaussian' (GG) beams \cite{av:generalized}, which provide a smooth interpolation of higher-order Gaussian profiles between linear, standing wave-like HG modes and rotating, OAM-carrying LG modes, naturally related to quantum angular momentum algebra \cite{da:analogies}.

Here, we consider a nondiffracting beam family counterpart to GG beams, interpolating between {\em standing} azimuthal Bessel beams and {\em travelling} plane waves with a transverse momentum component.
The family is based on the set of special function solutions of a well-understood physical problem, namely the eigenstates of a quantum pendulum \cite{condon:pendulum,grmc:mathieu} for a matter wave confined to a circle of radius $L$ subject to a constant gravitational field strength $g,$ which acts as the interpolating parameter for $0 \le g \le \infty.$
Such pendulum eigenstates are complex distributions on the circle with angle $\theta$ (with radius the pendulum length $L$), which we consider as the spectral distribution of our beams in Fourier space.

With a spectrum restricted to the circle (with $L \to k_{\mathrm{r}},$ the radial wavenumber), the propagating beam in real space has an invariant amplitude profile upon propagation \cite{durnin,mjgd:spread}.
The eigenfunctions of the quantum pendulum, first studied by Condon \cite{condon:pendulum} and subsequently re-discovered several times \cite{grmc:mathieu} are Mathieu functions \cite{grmc:mathieu,mclachlan}, which determine the spectra of our `pendulum beams'.
The ground state of the pendulum has been considered as a beam spectrum previously \cite{fas:uncertainty}, minimizing a certain angular uncertainty relation.
Our definition of this new beam family reveals interesting aspects of optical linear and orbital angular momentum, extends the GG beam interpolation to propagation-invariant beams, and provides an unusual appearance of the paraxial approximation. 
Furthermore, pendulum beams arise naturally as eigenfunctions of a certain operator which interpolates between angular and linear momentum.

Mathieu functions are more often encountered in optical beam physics in `Mathieu beams' \cite{MathieuBeams,lgmd:helical}, that is, the propagation-invariant beam family separated in elliptic coordinates $\xi,\eta$ with $x = f \cosh \xi \cos \eta, y = f \sinh \xi \sin \eta$ (analogous to the elliptic coordinate separation of the 2D Helmholtz equation \cite{grmc:mathieu,mclachlan}), with focal distance $f$.
With $f$ as a parameter, elliptic coordinates themselves interpolate between polar coordinates ($f = 0$) and cartesian coordinates (for which $f \to \infty$), so Mathieu beams interpolate between Bessel beams and plane waves.

As separable solutions of differential equations are usually represented as products of {\em real} functions in the separating coordinate system (thus having nodal lines in $x,y$ crossing at $90^{\circ}$), the Bessel beams appearing as the limit of Mathieu beams when $f\to 0$ are standing waves in azimuthal angle $\phi,$ with azimuthal dependence $\cos(\ell \phi), \sin(\ell \phi).$
Although a beam with standing azimuthal dependence $\cos(\ell \phi)$ is not an eigenfunction of the angular momentum operator $\hat{L} \equiv -\mathrm{i} \partial_{\phi},$ it {\em is} an eigenfunction of its square, $\hat{L}^2,$ with eigenvalue $\ell^2.$
A complex superposition of a symmetric (cosine-like) profile with $\pm\mathrm{i}$ times an antisymmetric (sine-like) profile gives a `helical' (i.e.~travelling azimuthal) beam: helical Mathieu beams \cite{lgmd:helical} limit to usual OAM-carrying Bessel beams, with $\mathrm{e}^{\mathrm{i}\ell \phi}$ azimuthal dependence.

These interpolations are analogous to the Gaussian mode interpolation between LG and HG modes by Ince-Gaussian (IG) modes \cite{Ince-Gaussian_beams} (solutions of the 2D harmonic oscillator separated in elliptic coordinates \cite{bkm:lie7}).
The HG modes are eigenstates of $\hat{H}_x - \hat{H}_y,$ i.e.~the difference between the harmonic oscillator hamiltonian operators in the $x$- and $y$-directions, and standing azimuthal LG modes are eigenfunctions of $\hat{L}^2.$
IG modes are eigenstates of the operator $\hat{L}^2 + f^2(\hat{H}_x - \hat{H}_y)$ in suitable units \cite{bkm:lie7}, where $f$ is again the focal distance. 
This contrasts with the GG modes, which are eigenstates of $\hat{L}\cos\alpha + (\hat{H}_x - \hat{H}_y)\sin\alpha$ for real angle $\alpha$ \cite{bkm:lie7}.

We now consider pendulum beams, starting by reviewing the solution of the quantum pendulum.
A matter wave of mass $M,$ confined to a circle of radius $L$ with angle $\theta$ to the vertical $y,$ has potential energy $M g y = M g L \cos\theta.$
It satisfies the time-independent Schr\"odinger equation $[-\hbar^2/(2ML^2)\partial_{\theta}^2 + M g L \cos\theta]\Psi(\theta) = E\Psi(\theta),$ where $\Psi(\theta)$ is the periodic eigenfunction $\Psi(0) = \Psi(2\pi)$ corresponding to the energy eigenvalue $E.$
With the substitutions $\chi = \theta/2, q = 4 M^2 L^3 g/\hbar^2, \alpha = 8M L^2 E/\hbar^2,$ and considering now $Y(\chi) = \Psi(\theta),$ the equation becomes
\begin{equation}
   Y''(\chi) + [\alpha - 2 q \cos(2\chi) ]Y(\chi) = 0,
   \label{eq:mathieueq}
\end{equation}
which is the standard form of Mathieu's equation \cite{grmc:mathieu,mclachlan,dlmf} with characteristic (eigenvalue) $\alpha$ and parameter (field strength) $q.$
In terms of the variable $\chi,$ $Y(0) = Y(\pi),$ meaning that only solutions of Eq.~(1) with period $\pi$ correspond to eigenfunctions of the pendulum.
These are of two types: Mathieu functions of even order which are even (odd) around $\theta, \chi = 0,$ $\mathrm{ce}_{2n}(\chi;q)$ [$\mathrm{se}_{2n}(\chi;q)$], with  characteristic values $a_{2n}(q)$ [$b_{2n}(q)$], where $n$ is an integer $n = 0, 1, 2, ..$ (1, 2, ...),  and we follow the notation conventions of \cite{mclachlan}.
The odd-order Mathieu functions, with antiperiod $\pi,$ do not appear as solutions of the quantum pendulum, but will appear later.

\begin{figure}
  \begin{center}
   \includegraphics[width=\linewidth]{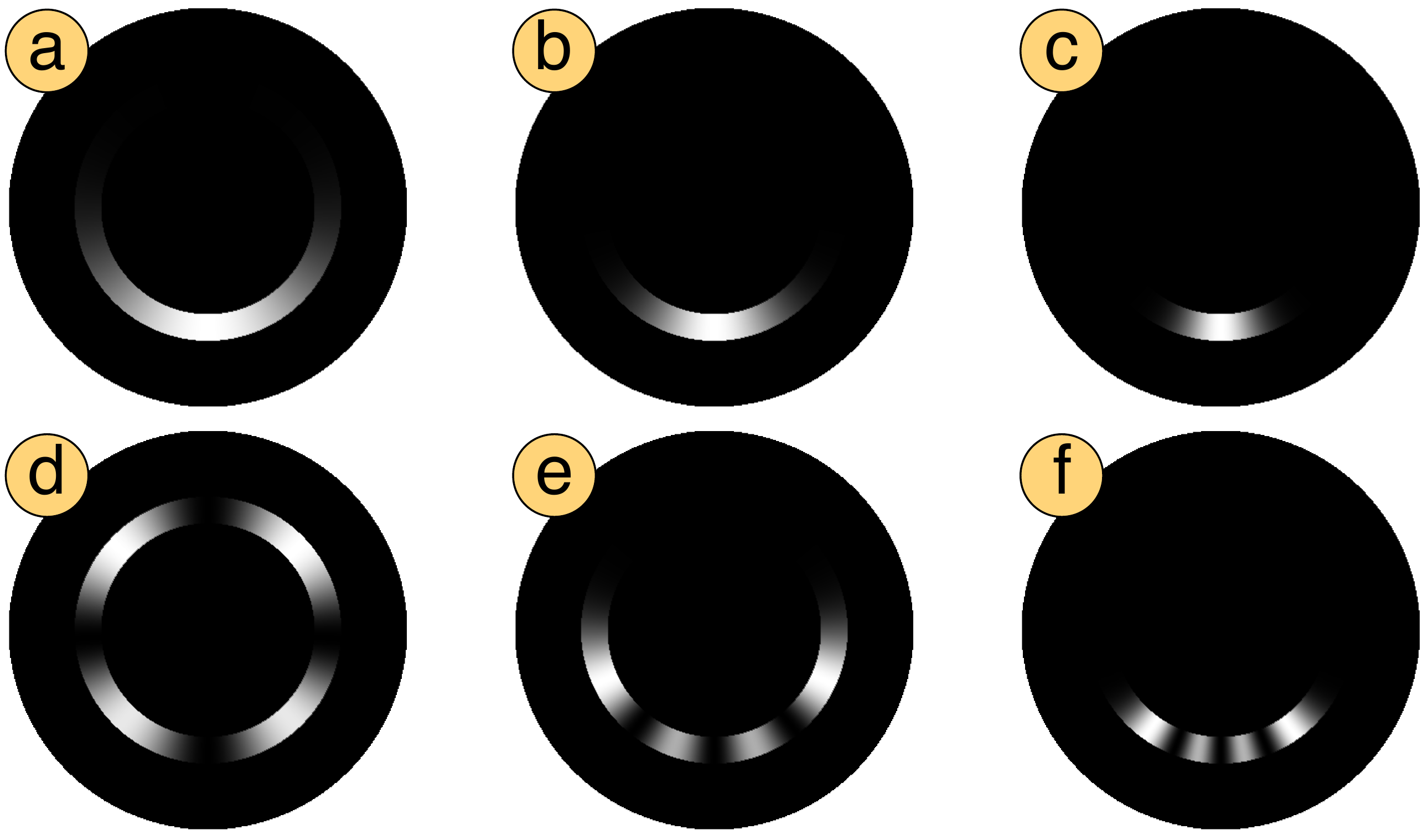}
   \caption{
   Quantum pendulum eigenfunctions, given by Mathieu functions in azimuth as in the text.
   (a)-(c) $n$ = 0, (d)-(f) $n = 4,$ (a), (d) $q = 5,$ (b), (e) $q = 50$, (c), (f) $q = 500.$
   These are the spectra (with radius $k_{\mathrm{r}}$) of pendulum beams.
   }
   \label{fig:pendulumefs}
 \end{center}  
\end{figure}

The pendulum eigenfunctions $\mathrm{ce}_0(\theta/2;q),$ $\mathrm{se}_4(\theta/2;q)$ are shown in Fig.~\ref{fig:pendulumefs}, for different choices of the parameter $q.$ 
When $g = q = 0,$ the wave is free to rotate on the circle with hamiltonian proportional to $\hat{L}^2,$ and characteristics $(2n)^2;$ when $n \neq 0,$ these are doubly degenerate with eigenfunctions $\exp(\pm\mathrm{i}n\theta)$ or $\cos(n\theta), \sin(n\theta).$
When $q > 0,$ the rotational symmetry is broken (selecting the $\cos,\sin$ states), with the characteristics $(2n)^2$ splitting to $a_{2n}(q), b_{2n}(q).$
When $q \gg 1,$ the low-$n$ states are confined to the region around the minimum potential $\theta = \pi;$ by the quantum correspondence principle, we expect these to be related to the harmonic oscillator eigenfunctions.

We define pendulum beams to be optical beams with these distributions as their transverse Fourier space spectra, where the ring radius $L$ is taken to be a fixed radial wavenumber $k_{\mathrm{r}}.$
Therefore their transverse Fourier representation is $\mathrm{ce}_{2n}(\theta/2;q)\delta(|\boldsymbol{k}|-k_{\mathrm{r}})$ for $n = 0, 1, 2, ...$ and $\mathrm{se}_{2n}(\theta/2;q)\delta(|\boldsymbol{k}|-k_{\mathrm{r}})$ for $n = 1, 2, ... .$
Since the spectrum has a fixed radial wavenumber, the beam also has a fixed longitudinal wavenumber $k_z$ and hence is propagation-invariant (and indeed, the transverse distributions solve the 2D Helmholtz equation) \cite{durnin,mjgd:spread}.
The real-space distribution is found by decomposing the pendulum Mathieu spectrum into Fourier components (in $\theta$), and then inverse Hankel transforming each component into real space $\boldsymbol{r} = (x,y) = (r \cos\phi,r\sin\phi)$ by 
\begin{equation}
   \delta(|\boldsymbol{k}| - k_{\mathrm{r}}) \cos(m \theta) \to \mathrm{i}^m \cos\!\left(m [\phi-\pi/2]\right) J_{m}(k_{\mathrm{r}} r),
   \label{eq:hankel}
\end{equation} 
and similarly for $\sin;$ the $\pi/2$ in the argument of $\cos$ is because we measure the real space azimuth with respect to $+x,$ although $\theta$ is the angle with respect to $+y.$

\begin{figure}
  \begin{center}
   \includegraphics[width=\linewidth]{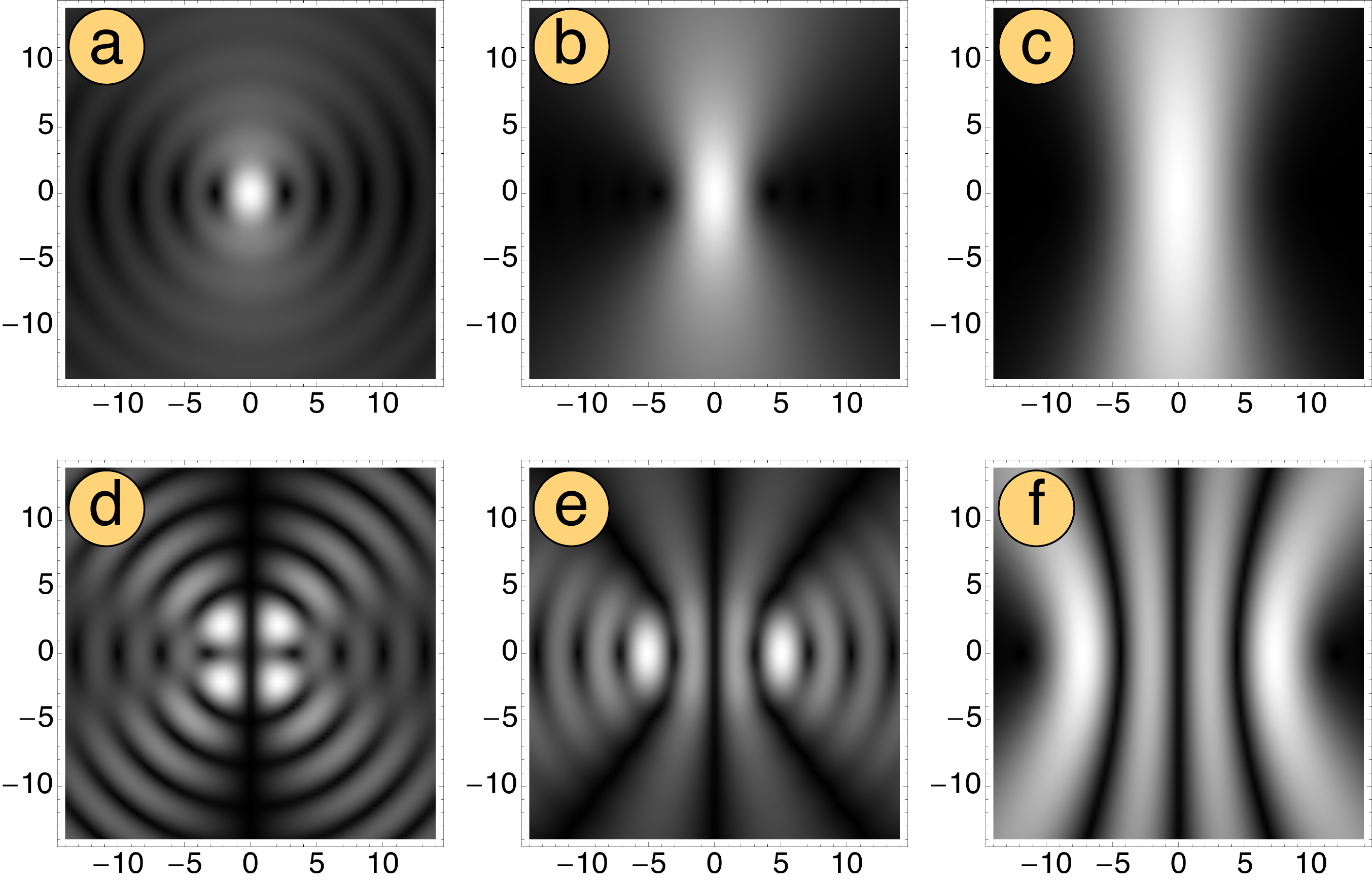}
   \caption{
   Pendulum beam intensities, with lengthscale $k_{\mathrm{r}}^{-1}.$
   Frames are Fourier transforms of those in Fig.~(\ref{fig:pendulumefs}): (a)-(c) $\Lambda_0,$ (d)-(f) $\Lambda_3,$ (a), (d) $q = 5,$ (b), (e) $q = 50$, (c), (f) $q = 500.$
   The profiles are propagation invariant, and interpolate between real Bessel beams ($q = 0$) and `beam-beams' ($q \gg 1$), whose profiles resemble propagating 1D HG wavepackets.
   }
   \label{fig:pendulumbeams}
 \end{center}  
\end{figure}

Thus pendulum beams $\Lambda_n(\boldsymbol{r};q),$ can be written in terms of superpositions of Bessel beams explicitly (keeping $z = 0$ for brevity) as follows:
\begin{eqnarray}
\!\!\!\!   \Lambda_{2n}(\boldsymbol{r};q)  & = & \!\! \sum_{m = 0}^{\infty} \mathrm{i}^m A_{2m}^{2n}(q) \cos\!\left(m \left[\phi-\tfrac{\pi}{2}\right]\right) J_{m}(k_{\mathrm{r}} r), \label{eq:pendulumeven} \\
\!\!\!\!   \Lambda_{2n-1}(\boldsymbol{r};q)  & = & \!\! \sum_{m = 0}^{\infty} \mathrm{i}^m B_{2m}^{2n}(q) \sin\!\left(m \left[\phi-\tfrac{\pi}{2}\right]\right) J_{m}(k_{\mathrm{r}} r), \label{eq:pendulumodd}
\end{eqnarray}
where the Fourier coefficients $A_{2m}^{2n}(q)$ [of $\mathrm{ce}_{2n}(\chi;q)$], and $B_{2m}^{2n}(q)$ [of $\mathrm{se}_{2n}(\chi;q)$], are defined by recursion relations \cite{mclachlan} section 3.10, \cite{dlmf} Eqs.~(28.4).
The reason for defining the orders of pendulum beams $\Lambda_{2n}, \Lambda_{2n-1}$ in terms of Mathieu coefficients as here follows naturally from the $q\gg 1$ regime, discussed below; unless $q = 0,$ $\mathrm{se}_0$ is not defined.
Examples of the pendulum beams $\Lambda_0(\boldsymbol{r};q)$ and $\Lambda_3(\boldsymbol{r};q)$ for the different $q$ of Fig.~\ref{fig:pendulumefs} are shown in Fig.~\ref{fig:pendulumbeams}.

When $q$ is small, pendulum beams have the intensity petals characteristic of a standing azimuthal Bessel beam.
As $q$ increases, the circular symmetry is broken.
When $q \gg 1,$ the transverse amplitude resembles a 1-dimensional Gaussian wavepacket (in $x$), propagating in the $-y$-direction, converging at an apparent `focus' at $y = 0.$
Despite appearing to be a propagating Gaussian wavepacket, this is the propagation-invariant transverse profile of a pendulum beam when $q$ is large; we will refer to beams in this regime as `Gaussian beam-beams'.

Unlike other beam families, pendulum beams have to be represented in terms of sums, such as by Eqs.~(\ref{eq:pendulumeven}), (\ref{eq:pendulumodd}); they cannot directly be expressed as products of elementary functions.
This is similar to GG modes, which are naturally expressed as sums of HG or LG modes with coefficients given by Wigner $D$-functions \cite{mrd:rows} (or equivalently Jacobi polynomials \cite{av:generalized}).
It is therefore more difficult to make general statements about how pendulum beams behave in different regimes.
We will now characterize pendulum beams when $q$ is small and large.

\begin{figure}
  \begin{center}
   \includegraphics[width=\linewidth]{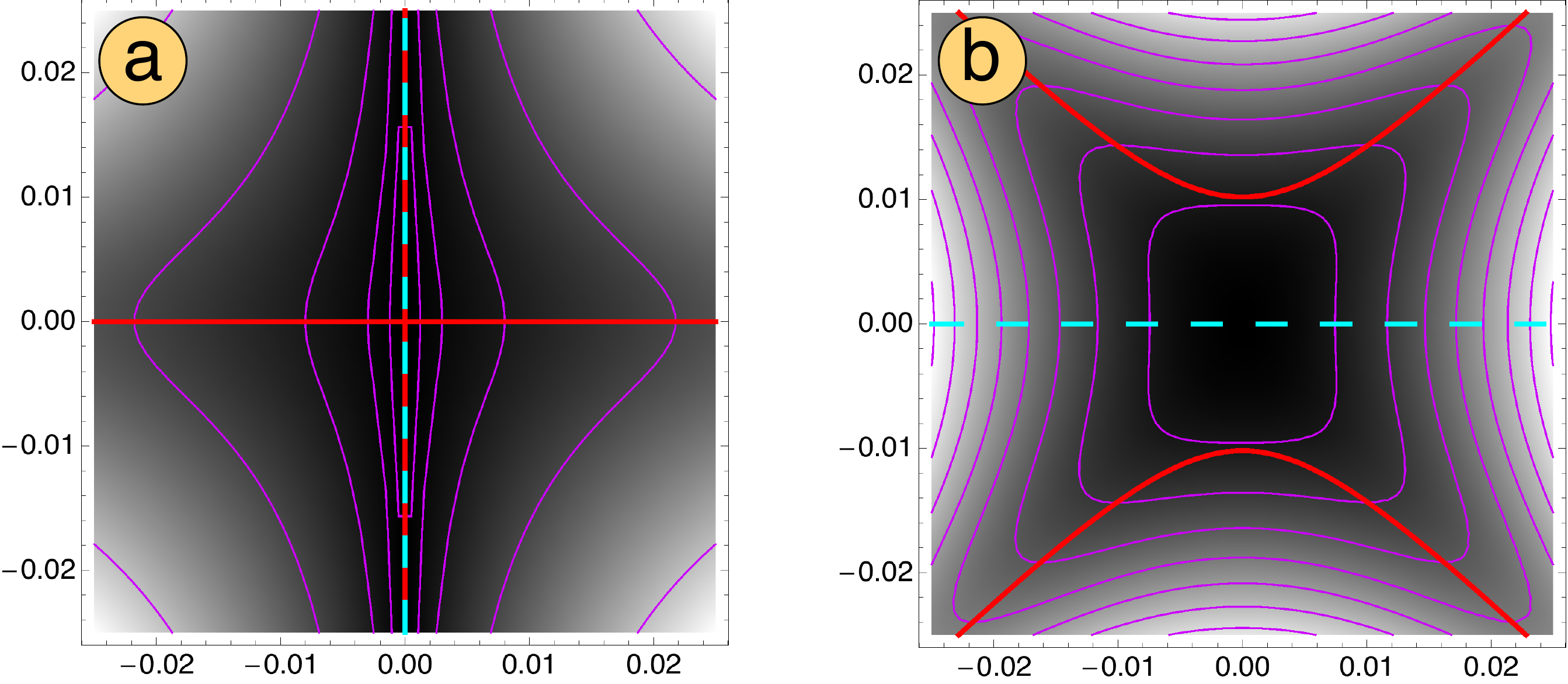}
   \caption{
   Amplitude distributions near the axis for $q \approx 0.$ 
   Thin lines are intensity contours, thick (dashed) lines real (imaginary) contours.
   (a) $n = 3,$ with amplitude $\propto 48 (k_{\mathbf{r}}^2 x y/q^2) - 8\mathrm{i}(k_{\mathbf{r}} x/q).$ 
   (b) $n = 4$ with amplitude $\propto 1 + 24(k_{\mathbf{r}} x/q)^2 - 24(k_{\mathbf{r}} y/q)^2 + 8\mathrm{i}(k_{\mathbf{r}} y/q).$
   The functions globally behave like $x y,$ and $x^2-y^2$ but the nongeneric axial singularity is perturbed by small $q,$ given by Eqs.~(\ref{eq:singuleven}), (\ref{eq:singulodd}).
   }
   \label{fig:localsmallq}
 \end{center}  
\end{figure}

When $0 < q \ll 1,$ the beam globally resembles the $q = 0$ Bessel beam.
However, when $n \neq 0,$ the region close to the beam axis with amplitude $\propto r^{n} \cos(n \phi)$ or $r^{n} \sin(n \phi)$ is most sensitive to the change in $q.$
Following singularimetric arguments analogous to those in \cite{mrd:rows,DennisGoette:PRL109:2012}, when $q$ and $k_{\mathrm{r}} r$ are comparably small, the neighborhood of the beam axis are given by finite series
\begin{eqnarray}
\!\!\!\!   \Lambda_{2n}(r,\phi) \appropto 
   \sum_{t = 0}^n \frac{(n+t-1)!}{(n-t)!t!} \left(\frac{2 \mathrm{i} k_{\mathrm{r}}r}{q}\right)^{t} \cos\!\left(t\left[\phi-\tfrac{\pi}{2}\right]\right),
   \label{eq:singuleven} 
   \\
\!\!\!\!   \Lambda_{2n-1}(r,\phi) \appropto 
   \sum_{t = 1}^n \frac{(n+t-1)!}{(n-t)!t!} \left(\frac{2 \mathrm{i} k_{\mathrm{r}}r}{q}\right)^{t} \sin\!\left(t\left[\phi-\tfrac{\pi}{2}\right]\right).
   \label{eq:singulodd} 
\end{eqnarray}
In \cite{mrd:rows,DennisGoette:PRL109:2012}, such finite series represent the breakup of a high-order optical vortex into $n$ single-charge vortices approximated by an analytic polynomial in $x + \mathrm{i} y.$ 
Here, we are considering the complex perturbation of a real amplitude distribution, which is harder to characterize mathematically; the high-order discrete rotational symmetry of the standing Bessel beam is broken, with a net phase increase in the $-y$-direction, as shown in Fig.~\ref{fig:localsmallq}.

In the regime $q \gg 1,$ the low-energy pendulum eigenstates are well approximated by harmonic oscillator eigenstates \cite{condon:pendulum}.
This is given by the leading order asymptotic approximation (given in \cite{dlmf} section 28.8), for which $a_n(q), b_{n+1}(q) \sim -2q,$ valid for $0 \le \chi < \pi,$
\begin{eqnarray}
   & & \!\!\! \mathrm{ce}_{m}(\chi;q),\, \mathrm{se}_{m+1}(\chi;q) \label{eq:largeq} \\
   & &  \; \sim \left[\frac{\pi\sqrt{q}}{2^{2m+1}(m!)^2}\right]^{1/4} \mathrm{e}^{-(q^{1/4}\cos\chi)^2}H_m\left(\sqrt{2}q^{1/4} \cos\chi\right),
   \nonumber
\end{eqnarray}
where $H_m$ is a Hermite polynomial.
Beams with spectra analogous to the right hand side of Eq.~(\ref{eq:largeq}) were considered in \cite{la:orthonormal}.

When $q$ is large, the pendulum spectrum is therefore only nonzero close to $\theta = \pi.$ 
In this region, the spectrum of $\Lambda_m(\boldsymbol{r};q)$ is well approximated by a 1D Hermite-Gaussian distribution of order $m$ in $k_x = -k_{\mathrm{r}}\sin\theta,$ with waist width $\sqrt{2}k_{\mathrm{r}}/q^{1/4}.$
This is analogous to the paraxial approximation, but in the transverse spectrum (as $k_z$ is fixed).
Being nondiffracting, the spectrum is confined to the circle of radius $k_{\mathrm{r}}$, but now in a sufficiently tight angle around $\theta = \pi$ that the circle may be approximated by a parabola.
Furthermore, the spectrum is Hermite-Gaussian in $k_x.$
Therefore, in real space, the transverse beam appears to be a paraxial 1D Hermite-Gaussian profile in $x$ propagating in $-y$ with focal width $q^{1/4}/\sqrt{2}k_{\mathrm{r}},$ as in Fig.~\ref{fig:pendulumbeams}(c),(f).
This implies the Gaussian-like profile in $x$ has a Rayleigh distance in $y$ of $\sqrt{q}/4k_{\mathrm{r}}.$

As $q \to \infty,$ the effective width in $x$ of this transversely propagating `Gaussian beam-beam' increases, limiting to a true plane wave propagating in $-y$ when $m = 0.$
When $m \neq 0,$ one finds superpositions of `polynomial beams' (differentiated plane waves) \cite{dgkma:polynomial}.

The Mathieu approximation above reveals another appealing feature of pendulum beams in the large $q$ regime.
By Eq.~(\ref{eq:largeq}), we have for $q \gg 1,$
\begin{equation}
   \mathrm{ce}_{2n}\left(\tfrac{\theta}{2},q\right) \approx \tfrac{1}{\sqrt{2}}\left[ \mathrm{ce}_{2n}\left(\theta-\tfrac{\pi}{2};\tfrac{q}{16}\right) + \mathrm{se}_{2n-1}\left(\theta-\tfrac{\pi}{2};\tfrac{q}{16}\right) \right],
   \label{eq:mathieudoubleapprox}
\end{equation}
that is, the (even) Mathieu function with argument $\theta/2$ and parameter $q$ can be expressed as the sum of adjacent even and odd Mathieu functions with shifted argument $\theta - \pi/2$ and parameter $q/16,$ and an analogous expression holds for $\mathrm{se}_{2n}(\theta/2,q)$.

This representation is informative as the Mathieu functions on the right hand side of Eq.~(\ref{eq:mathieudoubleapprox}) are in fact the Fourier representations of {\em Mathieu beams} \cite{MathieuBeams,lgmd:helical}.
This implies the following approximation for large $q$, 
\begin{equation}
   \Lambda_{2n}(\boldsymbol{r};q) \appropto \mathrm{Ce}_{2n}(\xi;Q)\mathrm{ce}_{2n}(\eta;Q) +\mathrm{i}\mathrm{Se}_{2n+1}(\xi;Q)\mathrm{se}_{2n+1}(\eta;Q),
   \label{eq:mathieubeamapprox}
\end{equation}   
where $Q = q/16 = f^2 k_{\mathrm{r}}^2/4$ and $\mathrm{Ce}, \mathrm{Se}$ denote modified Mathieu functions \cite{mclachlan,dlmf}. 
An analogous equation for odd order $\Lambda_{2n-1}$ also holds.
The transversely propagating beam-beam ($q \gg 1$) may thus be represented as a complex combination of even and odd (real) Mathieu beams of order differing by 1, just as an OAM-carrying Bessel beam is a complex combination of even and odd standing azimuthal Bessel beams of the same order.

The approximation of the hyperbolic pattern of a focused 1D Gaussian wavepacket by Mathieu functions has been made previously \cite{rc:exact}.
However, the present discussion illustrates how this representation naturally arises in a natural interpolation between Bessel beams, associated with OAM around the beam axis, and `beam-beams', associated with transverse linear momentum.
Just as OAM relies on a complex superposition of beams with $\cos(n \phi)$ and $\sin(n \phi)$ dependence, travelling transverse Gaussian beams approximately rely on a complex superposition of even and odd Mathieu functions.
However, when $q = 0,$ $\cos(n \phi)$ and $\sin(n \phi)$ must be combined to give OAM eigenstates, whereas when $q \to \infty,$ combinations of Mathieu beams of order $2n$ and $2n-1$ combine to give transverse linear momentum.
(This is mirrored in the Mathieu characteristics, for which, when $q\ll1,$ $a_n(q) \approx b_n(q),$ but $a_n(q) \approx b_{n+1}(q)$ for $q \gg 1$.)

In terms of operators, Mathieu beams are eigenfunctions of the `Mathieu operator' $\hat{L}^2 + \tfrac{2q}{k^2_{\mathrm{r}}}(\hat{p}_x^2 - \hat{p}_y^2);$ if $Y(\theta)$ is an eigenfunction of this operator with eigenvalue $\alpha$ and azimuth $\theta (=\chi),$ the operator equation in Fourier space can easily be arranged to give Mathieu's equation (\ref{eq:mathieueq}).
The preceding discussions imply that the pendulum beams are eigenfunctions of a related `Condon operator' $\hat{L}^2 + \tfrac{q}{2k_{\mathrm{r}}} \hat{p}_y,$ now with eigenvalues $\alpha/4,$ which interpolates between the squared angular momentum operator and the linear momentum operator, making the link with Mathieu beams natural, as well as the analogy with operators for different Gaussian beam families.

For experimental realizations with a hologram in the Fourier plane, only light on a circle of radius $k_{\mathrm{r}}$ can be transmitted \cite{mjgd:spread}; when $q \gg 1$ this is further concentrated on a small arc, which can be approximated by a parabola.
In fact, the recently-studied `Pearcey beams' \cite{Pearcey_beam} also have a spectrum restricted to a parabola, and in their focal plane they display a pattern reminiscent of the focusing of a Gaussian wavepacket (whereas in other planes this focusing appears aberrated).
Our consideration of operators suggests further insights into the paraxial approximation may be gained by considering beams which are eigenfunctions of the linear momentum operator singularly perturbed by the squared angular momentum operator, which, as we have seen, gives the `beam-beam' regime of pendulum beams.

\end{document}